\documentclass[aps,prl,amsmath,amssymb,twocolumn,longbibliography]{revtex4-2}
\usepackage[cp1251]{inputenc} 
\usepackage{amsmath}
\usepackage{amssymb}
\usepackage{bm}
\usepackage{datetime}
\usepackage[normalem]{ulem}
\usepackage{graphicx,color}
\usepackage{graphicx,scrtime}
\usepackage{tcolorbox}
\usepackage{soul}

\newcommand{\kk}{{\bm k}}

\newcommand{\uu}{{\bm u}}

\newcommand{\rr}{{\bm r}}

\newcommand{\be}{\begin{equation}}
\newcommand{\ee}{\end{equation}}
\newcommand{\ba}{\begin{eqnarray}}
\newcommand{\ea}{\end{eqnarray}}
\newcommand{\bse}{\begin{subequations}}
\newcommand{\ese}{\end{subequations}}
\newcommand{\beq}{\begin{eqnarray*}}
\newcommand{\eeq}{\end{eqnarray*}}

\newcommand{\im}{\mbox{Im}}
\newcommand{\ts}{\textstyle}
\newcommand{\scs}{\scriptstyle}

\newcommand{\eeta}{\bm\eta}

\newcommand{\sk}{

\vspace{.2cm}

}

\begin{document}
\title{%
Internal Stresses as Origin of the Anomalous Low-Temperature Specific Heat in Glasses}
\author{Walter Schirmacher}
\affiliation{Institut f\"ur Physik, Staudinger Weg 7, Universit\"at Mainz, D-55099 Mainz, Germany}
\affiliation{Center for Life Nano Science @Sapienza, Istituto Italiano di Tecnologia, 291 Viale Regina Elena, I-00161, Roma, Italy}
\author{Giancarlo Ruocco}
\affiliation{Dipartimento di Fisica, Sapienza Universit\`a di Roma, Piazzale Aldo Moro 2, 00185 Rome, Italy}
\affiliation{Center for Life Nano Science @Sapienza, Istituto Italiano di Tecnologia, 291 Viale Regina Elena, I-00161, Roma, Italy}

\begin{abstract}
	We apply a recently developed theory of the nonphononic vibrational  density of states (DOS)
in glasses to investigate the impact of local frozen-in stresses on the low-temperature specific heat.
Using a completely harmonic description we show that
the hybridization of the local nonphononic vibrational excitations with the waves leads to a low-frequency DOS,
in excess to the Debye one,
which varies linearly with frequency up to a certain crossover frequency, 
and then becomes constant. The actual value of the crossover depends of the ratio between the 
local stresses and the shear modulus. This excess DOS leads to a low-temperature specific heat 
with an apparent temperature exponent, which is between one and two, as observed experimentally. 
We discuss, how these findings may be utilized for the characterisation of glassy materials.
We further compare our findings,
which only rely on harmonic interactions,
with the predictions of other theories, which invoke anharmonic interactions and tunneling
for explaining the low-temperature behavior of the specific heat.

\end{abstract}
\maketitle
The low-frequency spectral properties of glasses
and the related low-temperature properties 
are very different from those of their crystalline counterparts.
\cite{zellerpohl71,phillips81,ramos22}.
For example, 
the specific heat of glasses does not
show the Debye temperature ($T$) law
$C(T) \sim T^3$, observed in crystals,
but, instead,
varies approximately as $C(T)\sim T^\alpha$ with an exponent
$\alpha$ between 1 and 2 \cite{zellerpohl71,lasjaunias74,lasjaunias75,stephens76,pohl81,liu93,phillips81,ramos22}.

A first interpretation of the very-low temperature thermal anomalies
of glasses 
was given by Anderson, Halperin, Varma and Phillipps in 1972
\cite{anderson72,phillips72}: It was conjectured that in a glass
bistable structural configurations would exist, characterized by a double-well
potential as a function of some configurational coordinate.
Further, tunneling between adjacent potential wells was
assumed to be possible, which
establishes quantum-mechanical two-level systems (TLS). It was
then assumed that
such TLS defect centers exist with a certain concentration and with
a broad (constant) distribution, $P(\Delta E)$, of energy splittings, $\Delta E$.
This TLS model leads to a specific heat varying linearly with
temperature ($\alpha=1$).

Although the tunneling model is nowadays widely accepted as explanation
of the low-temperature anomalies \cite{wurger97}, and is even called
``standard tunneling model'' \cite{klinger10}, severe criticism
has been expressed \cite{yuleggett88}. Indeed, in the temperature
range of 1 K the thermal de-Broglie wavelength of matter waves
 $\lambda=h[m k_BT]^{-1/2}$ (where $h$ and $k_B$ 
 are the Planck and Boltzmann constants, and
 $m$ is the mass of a molecular unit)
 is only a few percent of an angstrom
 for atomic masses larger than 10, which makes the existence of
 quantum tunneling
 in this temperature regime rather improbable. It has, however, been
 pointed out that bistable structural rearrangements might involve
 rather small
 single-atom displacements, which would then allow for
 quantum tunneling
 \cite{vegge01,lubchenko01}.

 Alternative suggestions for explaining the low-temperature thermal anomalies
 of glasses are based on interacting defects \cite{yuleggett88}, on
 elastic dipoles \cite{grannan90}, or on the soft-potential model, 
which  relies on the assumption of anharmonic defect states
\cite{karpov83,buchenau91,gurevich03}.
Most of the presently existing specific heat data
 have been evaluated using the TLS and soft-potential models
 \cite{ramos20,ramos22}.

As mentioned, the existing explanations of the anomalous low-T behavior of the
specific heat rely mainly on the presence of {\it anharmonic} interactions,
often in the form of double-well shaped interatomic potentials.
Here, we propose another possibility of explaining
 the low-temperature anomalies of the specific heat, which invokes neither quantum mechanics, nor anharmonicity. We
 present arguments that the spectral density of harmonic vibrations
 (density of states, DOS) $g(\omega)$ contains a low-frequency contribution,
 in addition to the Debye $g(\omega) \sim \omega^2$ law 
  which leads to a specific heat with a scaling exponent $\alpha$ which turns from 1 to 2 on lowering the temperature.
 These contributions
 arise from vortex-shaped pattern of vibrations around local
 frozen-in stresses.

 The presence of
frozen-in stresses in glasses is
 well known to glass blowers and is widely
 used for industrial applications, e.g.
 tailoring the surface properties of utility materials \cite{fotheringham21}.
Evidence for the (unavoidable) existence of local, randomly distributed stresses in
 glasses has been presented  in the literature 
 from computer simulations \cite{egami80,egami87,egami88}
 and from considering the microscopic derivation of
 continuum elasticity \cite{lutsko88,lutsko89,alexander98,schirmacher24}.
Experimentally internal stresses may be measured by indentation \cite{indentation}
 or by x-ray diffraction \cite{wumetglas15}.

 In a recent study, based on an entirely harmonic analysis,
the continuum limit
 of the Hessian (dynamical) matrix of a glass was studied
 \cite{schirmacher24}.
It was found, 
 that the low-frequency nonphononic (not wavelike) vibrational excitations of
 small computer-simulated glasses are mainly governed
 by the local stresses. 
 The nonphononic character of the excitations was guaranteed
 by considering small enough systems, which do not allow for
 low-frequency waves.
 By combining theory with molecular-dynamics simulations,
these excitations (called ``type-II modes'')
 were identified as nonirrotational, vortexlike states
 \cite{schirmacher24}. The spectrum of these states were shown to
 be related to the distribution of small stresses.  Small stresses imply small forces, which occur at molecular separations,
where the interpaticle-interaction potential has small values of its first derivative. In numerical simulations this artificially
makes the low-frequency DOS very sensitive to the 
smoothed cutoff (tapering). A tapering,
 which guarantees continuity of the first two derivatives
 of the potential ($m=2$) was shown to result in a DOS
 scaling as $g(\omega)\sim \omega^4$, whereas a tapering
 with $m=\infty$ leads to $g(\omega)\sim \omega^3$. 
 In real systems, for potentials with a minimum (like a Lennard-Jones potential)
 a scaling according to $g(\omega)\sim \omega^5$ emerges.
 These results were obtained by considering the modification
 of the frequency-dependent shear modulus by the
 nonphononic excitations  in the {\it absence} of waves in the interesting small frequency region.
 \cite{schirmacher24}.

 In the following we turn our attention to real, macroscopic, glasses and
consider the influence of such vortexlike
vibrational states on the DOS, coexisting with wavelike excitations
in the same frequency range.
 We shall show that, in addition to the modification of the
 shear elasticity,
 there exists also a direct contribution to the low-frequency
 DOS, which just reflects the statistics of the local stresses.
 As already mentioned, 
 such small stresses arise
from the minimum of the intermolecular potentials,
and exhibit a flat distribution, resulting in a DOS,
which depends only weakly on frequency.


In Refs. \cite{alexander84,alexander98,schirmacher24} it is shown that the continuum
limit of the harmonic energy of a system, interacting via a
pair potential $\phi(r_{ij})$ does not only involve the usual
strain degrees of freedom, as considered in elasticity theory \cite{bornhuang54}
and in heterogeneous-elasticity theory \cite{schirm06,schirm07,schirm14}, but
also nonirrotational, vortexlike vibrational modes, which are associated 
with local stresses and are coupled via these stresses to the
elastic degrees of freedom. 

The vorticities are defined in terms of vortexlike vibrational
displacement fields $\uu_\ell(\rr)$
centered around a local stress at $\rr_\ell$ as
\be
\eeta_\ell=\frac{1}{2}\nabla\times \uu_\ell(\rr)
\ee

The vortexlike local displacements, therefore, may be schematically expressed as
\be
\uu_\ell(\rr)=-\tilde\rr_\ell \times \eeta_\ell(\tilde\rr_\ell)\, ,
\ee
with $\tilde\rr_\ell=\rr-\rr_\ell$.
The vorticiy fields $\eeta$ are supposed to vanish for
large values of $|\tilde\rr_\ell|$.

In a simplified description, that we are going to present here,
we treat the local stresses $\sigma_\ell$
and the vorticities
$\eta_\ell(\tilde\rr_\ell,t)=|\eeta_\ell(\tilde\rr_\ell,t)|$ as scalars.
Further, we do not treat the elastic constants $\mu$ and $K$ (shear and bulk
modulus)
as fluctuating quantities, as assumed in Refs. \cite{schirm06,schirm07,schirm14},
because the focus is here on the
influence of the local stresses. 
Consequently, in the present treatment, the longitudinal and transverse sound velocities
\be\label{elast}
v_L^2=\frac{1}{\rho_m}\big[K+\frac{4}{3}\mu\big]\qquad 
v_T^2=\frac{1}{\rho_m}\mu
\ee
are also not considered to vary spatially. In (\ref{elast}) $\rho_m$ is the mass
density.
Including such spatially fluctuating
elastic constants makes it possible to include the description of
vibrational anomalies at higher frequencies
(``boson peak'') 
\cite{schirm06,schirm07,marruzzo13,schirmacher24}.

The coupled equations of motion for the longitudinal (L) and
transverse (T) elastic fields
$\uu_{L,T}(\rr,t)$ and the vorticities $\eta_\ell(\tilde\rr_\ell,t)$ can be written
as \cite{schirmacher24}
\ba\label{eqmo1}
\rho_m\bigg(\ddot \uu_{L,T}(\rr,t)-{v_{L,T}^2}\nabla^2 \uu_{{L,}T}(\rr,t)\bigg)
&=&\sum_\ell{\gamma_{L,T}^\ell}\sigma_\ell\nabla \eta_\ell(\tilde\rr_\ell,t)\nonumber\\
&&\\
\zeta\ddot \eta_\ell(\tilde\rr_\ell,t)+\sigma_\ell\eta(\tilde\rr_\ell,t)
&=&{\sum_{\alpha=L,T}\gamma_\alpha^\ell}\sigma_\ell\nabla\cdot \uu_{{L,}T}(\rr,t)\nonumber
\ea
Here 
$\gamma_{L,T}^\ell$ are coupling coefficients of the elastic strains 
with the local defect-induced rotational vibrations
and
may be considered to be a measure for the number of molecules involved in the anomalous mode with label $\ell$.

$\zeta$ is
an average local moment-of-inertia density
\be\label{inertia}
\zeta=\frac{1}{4}\rho_m\langle r^2_\perp\rangle
\ee
where $\langle r^2_\perp\rangle$ is an average distance
of the excitations from the defect center.

{We emphasize here, that in deriving these equations
\cite{schirmacher24}, the only underlying assumptions are that
the glass is composed of pairwise potentials, that it is stable,
and that it is structurally
disordered.}

{The mutual coupling between the waves and the vortices
causes a renormalization of the
diagonal Green's functions of the 
vortices (label $\ell$) 
and the waves (label $\alpha = L,T$), see Appendix A:}
\be
\label{green1}
G_{\ell\ell}(\omega)=\frac{1}{-\omega^2+\frac{1}{\zeta}\sigma_\ell+\Delta_\ell(\omega)}
\ee
{
\be
\label{green2}
G_{\alpha\alpha}(k,\omega)=
\frac{1}{-\omega^2+k^2v_\alpha^2+\Delta_{\alpha\alpha}(k,\omega)}
\ee
}
The frequency variable $\omega^2$ is understood to include an infinitesimally
small positive imaginary part.
The self-energies {$\Delta_{\alpha\alpha}(k,\omega)$}
and $\Delta_\ell(\omega)$ 
describe the hybridization of the
waves and the local vibrational excitations:
{
	\be\label{deltal}
\Delta_\ell(\omega)=
\frac{\sigma_\ell^2}{\rho_m\zeta}
\sum_{\alpha=L,T}\left(\frac{\gamma_\alpha^\ell}
{v_\alpha}
\right)^2
\ee
\be
\Delta_{\alpha\alpha}(k,\omega)=
\frac{k^2}{\rho_m \zeta}
\sum_\ell
{\gamma_\alpha^\ell}^2
\sigma_\ell^2
G_{\ell\ell}^{(0)}(\omega)
=\frac{k^2}{\rho_m}\Delta M_\alpha(\omega)
\ee
}
{As pointed out in \cite{schirmacher24}, the
wave renormalization gives rise to a frequency dependence
of effective elastic constants defined by
\be
M_\alpha(\omega)=\rho_m v_\alpha^2+\Delta M_\alpha(\omega)\, ,
\ee
so that we have
\be
G_{\alpha\alpha}(k,\omega)=\frac{1}{-\omega^2+\frac{1}{\rho_m}M_\alpha(\omega)k^2}
\ee
The density of states is then given by
\ba\label{dos}
g(\omega)&=&\frac{2}{\pi N}\omega\im\left\{
	\bigg\langle
\sum_\kk G_{LL}(k,\omega)\right.\nonumber\\
&&\qquad\qquad\left.
+2G_{TT}(k,\omega)+G_{\ell\ell}(\omega)
	\bigg\rangle_{P(\sigma)}
	\right\}\nonumber\\
&=&
g_D(\omega)
+\Delta g_{\rm ind}(\omega)
+\Delta g_{\rm dir}(\omega)
\ea
\sk
Here $\langle\dots\rangle_{P(\sigma)}$ indicates an average over
the distribution density $P(\sigma)$ of the stresses,
$g_D(\omega) \propto \omega^2$ is the Debye DOS, 
and $\Delta g_{\rm ind}(\omega)$ and
$\Delta g_{\rm dir}(\omega)$ are the indirect and direct modifications
of the DOS.
The indirect contributions are proportional to the imaginary parts
of the frequency-dependent moduli \cite{schirmacher24} according to
\ba\label{ind1}
\Delta g_{\rm ind}(\omega)
&\propto&\omega\im\big\{\Delta M_L(\omega)+2\Delta M_T(\omega)\big\}\nonumber\\
&\propto&\omega^5
P(\sigma)
\bigg|_{\omega^2=\sigma/\zeta}
\ea
}
As $\Delta g_{\rm ind}(\omega)$ has a frequency dependence with a rather high power, {(even higher than that of the Rayleigh
$\omega^4$ contribution, expected
from fluctuating elastic constants \cite{schirm07},)}
it is not relevant for the low-temperature specific heat, and we discard
it from the further discussion.

The direct stress-induced contribution to the density of states is
\be\label{direct1}
\Delta g_{\rm dir}(\omega)=N_\eta
\left\langle
\frac{2\omega}{\pi}
\im\bigg\{G_{\ell\ell}(\omega)
\bigg\}
\right\rangle_{P(\sigma)}
\ee
{
Because the transverse sound velocity in glasses is usually much smaller
than the longitudinal one \cite{pan21}, the term $\alpha=T$ in expression (\ref{deltal})
for 
the self-energy
$\Delta_\ell(\omega)$ will be dominant, so that we drop the longitudinal
term. 
We further 
assume a uniform, average transverse coupling
$\gamma=\gamma_T^{\ell}$, so that we have
\be
\Delta_\ell(\omega)=\frac{1}{\zeta\mu}(\gamma\sigma_\ell)^2\, ,
\ee
The direct contribution of
the stress-induced terms then becomes (Appendix B)}
\be
\Delta g(\omega)_{\rm dir}={\frac{N_\eta}{N}}2\omega P(\sigma)\bigg|_{\omega^2=f(\sigma)}\, ,
\ee
{where $N_\eta$ is the number of stress-related defect states, and}
\be
f(\sigma)=\frac{1}{\zeta}\sigma[1+{\gamma^2}\sigma/\mu]\, .
\ee
\sk
In order to be specific, we now assume that the glassy material
is composed of atoms or molecules, which interact via a pair potential
$\phi(|\rr_i-\rr_j|)=\phi(r_{ij})$. The absolute values of the fluctuating
local
stresses are given by \cite{lutsko88,schirmacher24}
\be
\sigma_{ij}=\bigg|\frac{1}{\Omega}r_{ij}\phi'(r_{ij})\bigg|
\ee
Here $\Omega$ is a microscopic volume with the size of the order of an interatomic spacing.
Because the potential usually has a minimum at a 
distance $r_0$, the small values of the stresses
are due to distances $r_{ij}$ near this value. 
In this vicinity we can write
\be
\sigma_{ij}(r_{ij})
=\sigma_1\big[r_{ij}-r_0\big]
\ee
In the regime around the minimum, where the
stresses are very small,
we can therefore
estimate the distribution density as
\be
\lim_{\sigma_{ij}\rightarrow 0}
P(\sigma_{ij})=
\frac{1}{\big|
\frac{d\sigma_{ij}}{dr_{ij}}
\big|}
4\pi\rho r_{ij}^2g(r_{ij})
={\rm const.}\doteq P_1\, ,
\ee
where $\rho=N/V$ is the density of molecules, $N$ their number,
$V$ the sample volume, and $g(r)$ the radial pair distribution
function \cite{hansen86}.
For such a constant distribution
density of small local stresses,
the expression (\ref{direct1}) leads to
(Appendix B)
\be\label{dos1}
\Delta g_\eta(\omega)
=n_\eta
\frac{\omega}{
	\sqrt{
		1+\left(\frac{\ts\omega}{\ts\omega_0}\right)^2}
	}
\ee
with $n_\eta=2N_\eta P_1\zeta/{N}$ and
\be\label{e19}
\omega_0^2=\frac{1}{4\zeta{\gamma^2}}\mu=
\frac{1}{\langle r_\perp^2\rangle{\gamma^2}}v_T^2\, ,
\ee
where we inserted relation (\ref{inertia}) for the inertia density.
We may introduce the Debye frequency $\omega_D=k_Dv_D$
\cite{ashcroft76}
with the Debye velocity
\be\label{e20}
v_D^{-3}=\frac{1}{3}\left(v_L^{-3}+2v_T^{-3}
\right)
\approx \frac{2}{3}v_T^{-3}
\ee
and the Debye wavenumber
$
k_D=\frac{1}{a}\sqrt[3]{6\pi^2}$
with the  typical distance between molecules
\mbox{$a=\sqrt[3]{m/\rho_m}$,}
where $m$ is the mass of a molecule
(See \cite{pan21} for values of $k_D$ of a large
number of glasses).
Then we can write for $\omega_0$
\be\label{omega0}
\omega_0=\frac{1}{\sqrt[3]{9\pi^2}}
\frac{a}{\gamma\sqrt{\langle r_\perp^2\rangle}}\omega_D\approx 0.2
\frac{a}{\gamma\sqrt{\langle r_\perp^2\rangle}}\omega_D
\ee
The frequency regime we are interested here
is the range $\omega\approx 10^{-2}\omega_D$. 
So, if
the ratio $\omega/\omega_0$ be of the order of 1,
the quantity $\sqrt{\langle r_\perp^2\rangle}\gamma/a$
becomes of the order of 500. This is a reasonal number
if we recall that this quantity, 
wich determines $\omega_0$ is the extent of the defects
in units of the intermolecular distance times the coupling $\gamma$.

Eq. (\ref{dos1}), together with the definitions in 
 Eqs. (\ref{e19})-(\ref{omega0}), represents the main result of the present paper, 
 as it gives a quantitative expression for the excess of the DOS at 
 low frequency in glasses.

Before we discuss the implications of our findings for the specific
 heat, we address the vibrational spectrum of glasses, and in particular
 the question, why a low-frequency DOS, which varies almost linearly with frequency has never been reported in the literature.
 As a matter of fact, both, simulational and experimental determinations
 of the low-frequency harmonic DOS encounter severe problems. In molecular-dynamics
 simulations
 the low-frequency eigenvalues are dominated by the presence
 of spurious standing waves due to the application of periodic
 boundary conditions. In order to detect ``nonphononic'' vibrational
 excitations extremely small samples have been investigated,
 in which the standing waves are suppressed \cite{lernerbouchbinder21,parisi19,parisi21}. However, as mentioned earlier,
 it turned out \cite{schirmacher24} that the low-frequency DOS of such
 small samples are -- via the local stresses -- very sensitive to
 the smoothing (``tapering'') of the potential near the imposed cutoff,
 leading (possibly) to artifacts. 
 In simulations with larger samples $N \sim 10^6$ to $10^7$ \cite{wang2018low,mizuno18b}
 a Debye $\omega^2$ spectrum below the boson peak is observed.  As 
we expect our predicted subquadratic DOS contribution to be very small compared
to the Debye law, and the low-frequency data in these simulations
exhibit some scatter, we
think they are not in conflict with our model.

 On the other hand, the experimental determination
 of the low-frequency vibrational DOS of harmonic excitations suffers
 from the presence of the anharmonic interactions, which is known
 to dominate the low-frequency DOS as exhibited
by incoherent inelastic scattering data of
 glasses\cite{wuttke95,chumakov04}). Coherent inelastic neutron- x-ray
 and Raman scattering data do not directly monitor the DOS
 \cite{taraskin97,schmid08}, and the
 low-frequency part in the GHz regime is, again, obscured by anharmonic
 excitations \cite{buchenau84,foret96,vacher05}. In order to reveal
 the low-frequency linear behavior of the vibrational DOS, one would
 have to do incoherent scattering experiments at very low frequencies
{\em and} very low temperatures, in order to avoid
 the anharmonic contributions.

	At frequencies near one-tenth of the Debye frequency the mean-free paths
 of the waves in glasses approach the Ioffe-Regel limit, i.e. its value becomes
 comparable to their wavelength. In this frequency regime a peak in
 the DOS, divided by $\omega^2$ is observed, called the boson peak.
 This anomaly, which shows up in the specific heat
 as a peak in the quantity $C(T)/T^3$, led to a huge number
of experimental and numerical investigations of the vibrational spectrum
of glasses. 
 \footnote{See Ref. \cite{ramos22}, in particular the review article of the present
 authors for the large number of references on the boson-peak related vibrational anomalies
 of glasses}. In Refs. \cite{schirm06,marruzzo13} spatially fluctuating
 elastic moduli have been identified as the main reason for the boson-peak
 anomaly. Other authors \cite{karpov83,buchenau91} advocated defectlike
 states, due to soft anharmonic potentials as reason for the
 anomalies. Such ``quasilocalized'' excitations \cite{laird91} are
 discussed frequently in the recent literature
	 on vibrations in glasses \cite{vogel23,lernerbouchbinder23}.
We believe that the stress-related defects states, discussed
in Ref. \cite{schirmacher24} and in the present Letter, are also relevant
in the boson-peak spectral regime, in particular the indirect
contribution of Eq. (\ref{ind1}). This should be the subject
of further investigations.

Now we focus on the specific heat,  that can be calculated 
{from the DOS} as \cite{ashcroft76}
\be\label{heat}
C_V(T)\sim\int_0^\infty d\omega g(\omega)
(\beta\omega)^2
\frac{e^{\beta\omega}
	}{
		\left[e^{\beta\omega}-1\right]^2
	}
\ee
with $\beta=\hbar/k_BT$.

In order to perform realistic calculations we 
restore the  Debye DOS
\be
g_D(\omega)=\frac{3}{\omega_D^3}\omega^2\theta(\omega_D-\omega)
\ee
where $\theta(x)$ is the Heaviside step function.
We then write  the relevant
DOS as%
\footnote{Here we have considered that
$n_\eta$ is of the order of 10$^{-4}$ ps$^2$, so that
the non-Debye term does not contribute much to the DOS
in the frequency regime,
where the Debye term is large.
Therefore the
normalization of the DOS should not be affected by
the non-Debye term.}
\ba\label{dos}
g(\omega)&=&g_D(\omega)+\Delta g_\eta(\omega)\nonumber\\
&=&
\frac{\ts 3}{\omega_D^3}\omega^2\theta(\omega_D-\omega)+
n_\eta\bigg(\frac{\omega}{\scs\sqrt{{\ts 1+}\left({\ts\frac{\omega}{\omega_0}}\right)^2}}
\bigg)\, ,
\ea
 
The specific heat can be now calculated numerically by Eq.
(\ref{heat}), using (\ref{dos}). {Because the DOS asymptotically
becomes linear for $\omega\rightarrow 0$ we predict
$C(T)\propto T^2$ for very low temperatures. If $T$ becomes
comparable with $T_0=\hbar\omega_0/k_B$ a smoothing toward
a linear behavior is expected. Finally, depending on $n_\eta$,
at higher temperature,
Debye's $T^3$ is approached}
\footnote{We do not consider the boson-peak anomaly, which
occurs at higher frequencies/temperatures than of interest here.
}.

\begin{figure}
	\includegraphics[width=8cm]{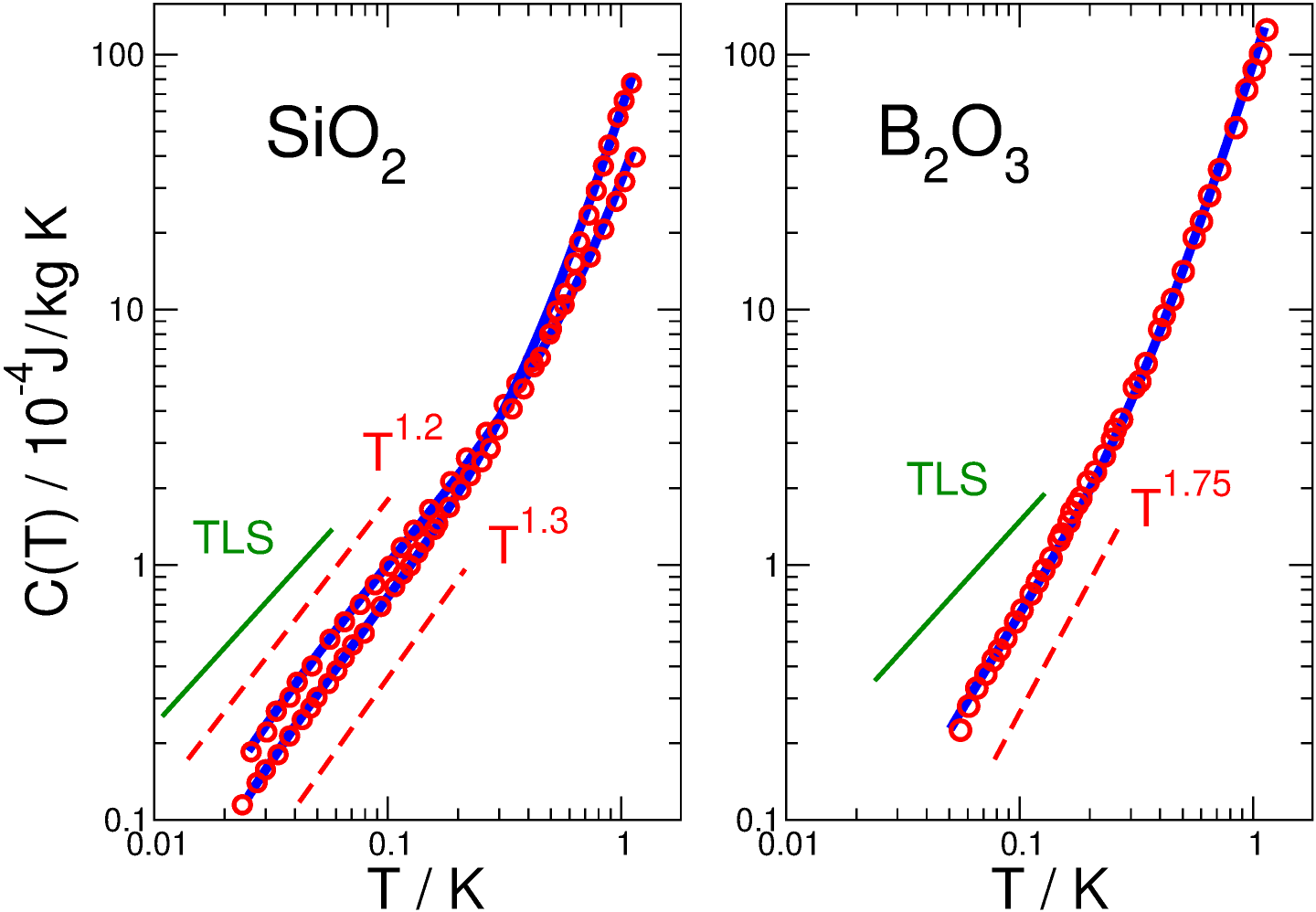}
	\caption{
Left:
Specific heat of SiO$_2$ 
	measured by Lasjaunias {\it et al.} \cite{lasjaunias75}
	together with fits according to Eqs. (\ref{heat}) and
	(\ref{dos}) (blue lines). The data set with apparent temperature
	dependence $C(T)\sim T^{-1.2}$ is Suprasil with
	$\approx $ 1 \% OH content, the data with $C(T)\sim T^{-1.3}$
	is Suprasil W with only 1 ppm OH. The fit parameters
	for Suprasil are 
	$\omega_0=4.5\cdot 10^{-3}$ ps$^{-1}$,
	$n_\eta=5.76\cdot 10^{-4}$ ps$^2$, for Suprasil W
	$\omega_0=6.25\cdot 10^{-3}$ ps$^{-1}$,
$n_\eta=1.28\cdot 10^{-4}$ ps$^2$.
\newline
Right:
Specific heat of B$_2$O$_3$, measured by Lasjaunias
	{\it et al.} \cite{lasjaunias74} 
(symbols) and
calculated via Eqs. (\ref{heat}) and
(\ref{dos}) 
(full blue line). We used the parameters
\mbox{$\omega_0$ = 1.5$\cdot 10^{-2}$ ps$^{-1}$,}
	$n_\eta=3.2\cdot 10^{-4}$ ps$^2$.\newline
The green lines labeled ``TLS'' indicate
a linearly varying specific heat, which is
predicted by the two-level-system tunneling model. The dashed red lines indicate the apparent slopes reported in the original papers  \cite{lasjaunias74,lasjaunias75}.
\\
For the Debye frequencies we use the values
	{$\Theta_D=342\mbox{ K }\Rightarrow
	\omega_D=\frac{k_B}{\hbar}\Theta_D$ = 44 ps$^{-1}$ 
	(SiO$_2$), $\Theta_D=153\mbox{ K }\Rightarrow
	\newline
	\omega_D$ = 20 ps$^{-1}$ 
	(B$_2$O$_3$) \cite{freeman86}.}
\label{heatex}
}
\end{figure}

In Fig. \ref{heatex} we show the low-temperature specific heat
of borate and silica glass as measured some time ago by
Lasjaunias {\it et al.} \cite{lasjaunias74,lasjaunias75}. All datasets
do not show a linear but a superlinear temperature dependence, which
is obviously well accounted for by our model. We indicated  by a green full line the linear
law $C(T)\sim T$ corresponding to the prediction of the TLS model
in the figure.

The values of $\omega_0$ obtained in the fits 
for SiO$_2$
correspond to
temperatures $\hbar\omega_0/k_B \approx $ 40 mK .
For B$_2$O$_3$ we get
\mbox{$\hbar\omega_0/k_B$ $\approx$ 800 mK}. 
Therefore, we
expect that the $C(T)$ of SiO$_2$ will bend down to show a $T^2$ law in
the $T$ range below 50 mK. $C(T)$ for B$_2$O$_3$ extends already
to $\hbar\omega_0/k_B$ and will also slightly bend down toward
$C(T)\sim T^2$.

Using Eq. (\ref{omega0}),
we may evaluate the parameter 
$\sqrt{\langle r_\perp^2\rangle}\gamma$,
the extent of the defects times the coupling
for SiO$_2$ and B$_2$O$_3$
from the values of $\omega_0$ obtained by fitting the
specific heat. Using the values of the Debye temperature
of these materials \cite{freeman86}
$\Theta_D$ = 342 K (SiO$_2$),
$\Theta_D$ = 153 K (B$_2$O$_3$), we get, according to
Eq. (\ref{omega0})
$\sqrt{\langle r_\perp^2\rangle}\gamma/a \sim$ 2000 for Suprasil, 
$\sqrt{\langle r_\perp^2\rangle}\gamma/a \sim$ 1440 for Suprasil W, and
$\sqrt{\langle r_\perp^2\rangle}\gamma/a \sim $ 50
for B$_2$O$_3$. The much smaller value of this parameter in B$_2$O$_3$
may be rationalized, taking into account that in this material, as opposed
to SiO$_2$, there is a tendency for the formation of layered structural motifs (boxol rings,
\cite{boroxol}),
which might
lead to defect states involving fewer molecular units.
\sk

For the low-temperature specific heat we envisage the following
scenario: The boson-peak anomaly is in most materials located
at one-tenth of the Debye temperature $\Theta_D=\hbar\omega_D/k_B$.
Below the boson peak,  the quantity $C(T)/T^3$  usually levels off and
then begins to rise with decreasing temperature, indicating the
cross-over toward the low-temperature regime we are talking about
\cite{ramos20,ramos22}
The specific heat in glasses is usually
measured down to $\approx 10^{-2} \Theta_D $.
According to our findings,
the temperature dependence in this regime is related to
the amount and spatial extension of frozen-in stresses in glasses.

It is important to underline that our model can be tested against future experiments, indeed its prediction implies that $C(T)$ at lower temperatures
(around 1 mk) should bend down toward a $T^2$ law.

It is also interesting to
study the development of the low-temperature specific heat as a function
of thermal, chemical or pressure treating. A corresponding change
in the prefactor and the crossover parameter $\omega_0$
will give information on
the salient features of the internal stresses. This information
on the internal stresses will add to the existing disorder classification
of glasses, provided by heterogeneous-elasticity theory
\cite{schirm06,schirm07,schirm14,pan21}.
In fact, in a recent study of ultrastable glasses
\cite{ramos23} it has been demonstrated that the low-temperature
non-Debye specific heat, which was present in the conventional
glasses and attributed to tunneling systems, was absent in
the ultrastable glasses. Within our new interpretation this 
means that the frozen-in stresses are strongly reduced in the ultrastable
glasses.
\sk
Let us finally mention that, of course, the contribution of fluctuating
elastic constants, which give rise to Rayleigh scattering and the
boson peak \cite{schirm06,schirm07,schirm14}, are not included in the
present study, which focuses on the frequency regime much below the
boson peak. A combined theory (generalized heterogeneous-elasticity theory)
for the DOS has been formulated recently \cite{schirmacher24}, 
and we shall apply 
it to the specific 
heat (in comparison with experimental data) in a forthcoming
paper.
\section{End Matter}
{
\subsection{Appendix A: Calculation of the Green's functions}
We start with the coupled equations (3) of the main text.
A Fourier transform with respect space and time gives
($\alpha=L,T$)
{\footnotesize
\ba\label{eqmo2}
\big[-\omega^2 +v^2_{\alpha}k^2\big]
u_{\alpha}(k,\omega)
&=&\frac{1}{\rho_m}\sum_\ell\gamma_{\alpha}^{\ell}\sigma_\ell
ik\,e^{i\kk\rr_\ell}\eta_\ell(k,\omega)\nonumber\\
&&\\
	\big[-\omega^2+
\frac{1}{\zeta}\sigma_\ell\big]
e^{i\kk\rr_\ell}\eta_\ell(k,\omega)
&=&\frac{1}{\zeta}\sum_{\alpha=L,T}\gamma^\ell_\alpha\sigma_\ell i\kk\cdot u_\alpha(k,\omega)\nonumber
\ea
}

We may eliminate the vortex fields from these coupled equations, which gives
the following effective
equations for the longitudinal and transverse fields
{\footnotesize
\be
\big[-\omega^2 +v^2_{\alpha}k^2\big]
u_\alpha(k,\omega)=
-\sum_{\alpha'=L,T}\Delta_{\alpha\alpha'}(k,\omega)u_{\alpha'}(k,\omega)
\ee
with
\be
\Delta_{\alpha\alpha'}(k,\omega)=
\frac{k^2}{\rho_m \zeta}
\sum_\ell 
\gamma_\alpha^\ell
\gamma_{\alpha'}^\ell
\sigma_\ell^2
G_{\ell\ell}^{(0)}(\omega)
\ee
}
We see that the coupling between the elastic degrees of freedom
and the vortices induces an indirect coupling between the
longitudinal and the transverse waves. Disregarding this coupling,
we obtain for the diagonal
Green's functions
\be\label{green1}
G_{\alpha\alpha}(k,\omega)=\frac{1}{
	-\omega^2 +v^2_\alpha k^2
+\Delta_{\alpha\alpha}(k,\omega)
}
\ee
We may as well eliminate the wave fields to obtain
\ba\label{eqmo4}
&&\big[-\omega^2+
\frac{1}{\zeta}\sigma_\ell
\big]
\eta_\ell(k,\omega)\nonumber\\
&=&-
e^{-i\kk\rr_\ell}
\frac{1}{\rho_m\zeta}
\sigma_\ell
\sum_\alpha
\gamma_\alpha^\ell
\frac{k^2}{-\omega^2+k^2v_\alpha^2} \nonumber\\
&&\qquad\times
\sum_{\ell'}\gamma_{\alpha}^{\ell '}\sigma_\ell e^{ik\rr_{\ell'}}
\eta_{\ell'}(k,\omega)
\ea
If we average over the positions $\rr_\ell$ we get
\be
\sigma_\ell\sigma_{\ell'}
\left\langle
e^{-ik\rr_\ell}
e^{ik\rr_{\ell'}}
\right\rangle_{\rr_\ell}
=\sigma_{\ell}^2\delta_{\ell\ell'}
\ee
from which we obtain 
\ba\label{eqmo4}
&&\big[-\omega^2+
\frac{1}{\zeta}\sigma_\ell
\big]
\eta_\ell(k,\omega)\nonumber\\
&=&-
\frac{1}{\rho_m\zeta}
\sum_\alpha
(\sigma_\ell\gamma_\alpha^\ell)^2
\frac{k^2}{-\omega^2+k^2v_\alpha^2} \nonumber\\
&&\qquad\times
\eta_{\ell}(k,\omega)
\ea
In the low-frequency limit this becomes
\be\label{eqmo5}
\big[-\omega^2+
\frac{1}{\zeta}\sigma_\ell
\big]
\eta_\ell(k,\omega)
=-\Delta_\ell(\omega)
\eta_\ell(k,\omega)
\ee
with 
\be
\Delta_\ell(\omega)=
\frac{\sigma_\ell^2}{\rho_m\zeta}
\sum_\alpha\left(\frac{\gamma_\alpha^\ell}
{v_\alpha}
\right)^2
\ee
The local Green's function of the vortices is then
obtained as
\be
G_{\ell\ell}(\omega)=\frac{1}{-\omega^2+
\frac{1}{\zeta}\sigma_\ell
+\Delta_\ell(\omega)
}
\ee
}
\subsection{Appendix B: Detailed derivation of the
density of states}

From Eq. 
(\ref{direct1}) 
we have
\be
\Delta g_\eta(\omega)
=N_\eta P_1\frac{2\omega}{\pi}
\int_0^{\sigma_{\rm max}}
\omega\,\delta\bigg(\omega^2-f(\sigma)\bigg)d\sigma
\ee
with the substitution $\lambda=\omega^2$ we get
the level density
\be\label{int}
\Delta \rho_\eta(\lambda)=\frac{1}{2\omega}g\bigg(\omega\!=\!\sqrt{\lambda}\bigg)
=N_\eta P_1\frac{1}{\pi}\int_0^{\sigma_{\rm max}}
\delta\bigg(\lambda-f(\sigma)\bigg)d\sigma
\ee
{
We have restricted the integration range in (\ref{int}) to postive
values of $\sigma$. Negative values would lead to negative values
of $\lambda=\omega^2$, which is inhibited by the stability requirement
for the spectrum.
}
We now further substitute
\be
f(\sigma)=\frac{1}{\zeta}\bigg(\sigma+\frac{{\gamma^2}}{\mu}\sigma^2\bigg)
\ee
from which follows
\be
df=d\sigma\frac{1}{\zeta}\bigg(1+\frac{2{\gamma^2}}{\mu}\sigma\bigg)
=d\sigma\frac{1}{\zeta}
\sqrt{
	1+\frac{4}{\mu}{\gamma^2}\zeta\lambda
}\, ,
\ee
where the second equality follows from the solution
of the equation $\lambda=f(\sigma)$ for $\sigma$.
We obtain
\be
\Delta \rho_\eta(\lambda)=
N_\eta P_1\frac{\zeta}{\pi}
\frac{1}{
	\sqrt{
		1+\frac{4}{\mu}{\gamma^2}\zeta\lambda
}}
\ee
from which follows
\be
\Delta g_\eta(\omega)=
N_\eta P_1\frac{\zeta}{\pi}
\frac{2\omega}{
	\sqrt{
		1+\frac{4}{\mu}{\gamma^2}\zeta\omega^2
}}
\ee

\end{document}